\documentclass[aps,pra]{revtex4}

\usepackage{epsfig,pstricks}
\usepackage{amsmath,amssymb}
\usepackage{color}

\def\CC{{\rm\kern.24em \vrule width.04em height1.46ex depth-.07ex
\kern-.30em C}}
\def\P{{\rm I\kern-.25em P}}
\def\NN{{\rm I\kern-.25em N}}
\def\RR{{\rm
         \vrule width.04em height1.58ex depth-.0ex
         \kern-.04em R}}
\def\id{{\rm 1\kern-.22em l}}
\def\ZZ{{\sf Z\kern-.44em Z}}
\def\tr{{\rm tr}\;}
\newtheorem{pdef}{Definition}[section]

\newenvironment{eqblock}[2]{\beq\label{#2}\begin{array}{#1}}{\end{array}
                                \eeq}
\newenvironment{neqblock}[1]{\[\begin{array}{#1}}{\end{array}\]}
\newcommand{\beqb}{\begin{eqblock}}
\newcommand{\eeqb}{\end{eqblock}} 
\newcommand{\nbeqb}{\begin{neqblock}}
\newcommand{\neeqb}{\end{neqblock}} 
\newcommand{\eps}{\varepsilon}
\newcommand{\beq}{\begin{equation}}
\newcommand{\beqa}{\begin{eqnarray}}
\newcommand{\eeq}{\end{equation}}
\newcommand{\eeqa}{\end{eqnarray}}
\newcommand{\nbeqa}{\begin{eqnarray*}}
\newcommand{\neeqa}{\end{eqnarray*}}
\newcommand{\bra}[1]{\langle #1 |}
\newcommand{\ket}[1]{| #1 \rangle}
\newcommand{\expect}[1]{\langle #1 \rangle}
\newcommand{\Matrix}[2]{\left( \begin{array}{#1} #2 \end{array}
  \right)}

\def\DJo{$\;$\kern-.4em \hbox{D\kern-.8em\raise.15ex\hbox{--}\kern.35em okovi\'c}}

\begin{document}

\title{SL-invariant entanglement measures in higher dimensions: the case of spin $1$ and $3/2$}
\author{Andreas Osterloh}
\affiliation{Institut f\"ur Theoretische Physik, 
         Universit\"at Duisburg-Essen, D-47048 Duisburg, Germany.}
\email{andreas.osterloh@uni-due.de}
\begin{abstract}
An SL-invariant extension of the concurrence to higher  
local Hilbert-space dimension
is due to its relation with the determinant of the matrix of
a $d\times d$ two qudits state, which is the
only SL-invariant of polynomial degree $d$. 
This determinant is written in terms of antilinear 
expectation values of the local $SL(d)$ operators.
We use the permutation invariance of the comb-condition for 
creating further local antilinear operators which 
are orthogonal to the original operator.
It means that the symmetric group acts transitively on the space of combs of a given order.
This extends the mechanism for writing $SL(2)$-invariants 
for qubits to qudits. I outline the method, that in principle 
works for arbitrary dimension $d$,
explicitly for spin 1, and spin 3/2. 
There is an odd-even discrepancy: whereas for half odd integer spin 
a situation similar to that observed for qubits is found, 
for integer spin the outcome is an asymmetric invariant of polynomial degree 
$2d$. 
\end{abstract}

\maketitle

\section{Introduction}
Since the development of quantum information theory, the importance of 
entanglement as a resource in physics has become clearer, 
and its quantification is hence an outstanding task.  
The minimal requirement for an entanglement measure is the symmetry under
local $SU$ operations~\cite{MONOTONES} and a lot of theoretical work has 
been devoted to its invariance group~\cite{LindenPopescu98,GrasslRoetteler98,CarteretLinden99,Wallach,Leifer04,Wallach05}. However, this 
group must be enlarged to the complexification of $SU$, the $SL$, 
which then encompasses the Stochastic Local Operations and Classical Communication 
(SLOCC). The general linear $GL$ group of general local operations admits 
in its closure also projective measurements.
The invariance group of $SL(2)$ has been explored in the mathematics 
and physics literature 
~\cite{Cayley,Hilbert,Wong00,Luque02,Brylinski02a,Jaeger03,JaegerII03,Teodorescu03,
Briand03,Levay05,LevayString06,Wallach05,Luque05,VerstraeteDM03,OS04,OS05,DoOs08}. 
It has several nice properties, such as 
it leads automatically to entanglement monotones,
it automatically contains the entangled states it measures~\cite{OS09}, 
and it determines a relation between the 
polynomial degree of an invariant that can possibly detect the 
state~\cite{DoOs08,OS09,JohannsonO13}. 
Furthermore, first results for the $SL(2)$ concerning optimal decompositions 
and entanglement of mixed states are also known
~\cite{Wootters98,Uhlmann00,LOSU,KENNLINIE,ViehmannII,EOSU,Eltschka2012,Siewert2012} 
such that we are close to a breakthrough towards measuring the 
invariants of $SL(2)$ experimentally. 
This astonishingly powerful tool of $SL$-invariance still awaits its extension
on the local operator level to general spin, though formally it is known 
(see for example \cite{VerstraeteDM03}).

Here, I will generalize the theory of local antilinear operators 
presented in~\cite{OS04,OS05} to higher 
spins $1$ and $3/2$~\footnote{Preliminary results have been reported in ~\cite{OS09}.}.

The paper is laid out as follows. In Section \ref{combs}, I review the idea of
combs for qubits, and devote the next Section \ref{higherSpin} to elaborate 
the combs for spin $1$, and $3/2$. 
The conclusions and an outlook are given in Section \ref{concls}.

\section{The concept of local anti-linear operators}\label{combs}
 
The fundamental concept that represents the basis for the construction method
of $SL(2)$-invariant operators, is that of a {\em comb}. It is a local antilinear operator
$A$ with zero expectation value for all states of the local Hilbert space
${\cal H}_i$\cite{OS04,OS09}. Here, I give a brief summary of this formalism.
A condition for an operator to be a comb is hence 
\beq\label{def:Toeter}
\bra{\psi} A_i \ket{\psi}=\bra{\psi} L_i {\cal C}\ket{\psi}
	=\bra{\psi} L_i \ket{\psi^*}\equiv 0\quad 
\forall\, \psi\,\in\, {\cal H}_i\; ,
\eeq
where $\cal C$ is the complex conjugation in the computational basis
\nbeqa
\ket{\psi^*}&:=&{\mathcal C}\ket{\psi}\equiv 
{\mathcal C}\sum_{j_1,\dots,j_q=0}^1 \psi_{j_1,\dots,j_q}\ket{j_1,\dots,j_q}\\
&=&\sum_{j_1,\dots,j_q=0}^1 \psi_{j_1,\dots,j_q}^*\ket{j_1,\dots,j_q}\; .
\neeqa 

In order to possibly vanish for every local state, the operator necessarily 
has to be antilinear 
(a linear operator with the above property is identically zero).
The problem is, to identify a comb that is
regular on the space it acts on, and with equal modulus of the corresponding eigenvalues 
due to the invariance property~\cite{OS09}.
We will call such an antilinear operator
$A^{(1)}: \mathfrak{h}\to\mathfrak{h}$ to be of {\em order 1}, where $\mathfrak{h}$ is the 
local Hilbert space.
The expectation value of $A^{(1)}$ is a homogeneous polynomial of lowest
possible degree $2$. 

In general we will call a (local) antilinear operator 
$A^{(n)}: \mathfrak{h}^{\otimes n}\to\mathfrak{h}^{\otimes n}$ to
be {\em of order n}; its expectation value is a homogeneous polynomial
of degree $2n$ in the coefficients of the state $\ket{\psi}$.
It is worthwhile noticing that the $n$-fold tensor product
$\mathfrak{h}^{\otimes n}$, a comb of order $n$ acts on, is made of
$n$-fold copies of one single qubit (or qudit) state. In order to distinguish
this merely technical introduction of a tensor product of copies of states 
from the physically motivated tensor product of different qubits
we will denote the first tensor product of copies with the symbol $\bullet$,
and hence write $A^{(n)}: \mathfrak{h}^{\bullet n}\to\mathfrak{h}^{\bullet n}$.
I will make sometimes use of the abbreviation
\beq
\bra{\psi} L_i C\ket{\psi}=:\expect{L_i}=:(\!( L_i )\!)\; .
\eeq
and will use the term comb also for the corresponding linear operator to a comb.

The conditions we are searching for, can then be written as
\beqa
 \bra{\psi} A^{(1)} \ket{\psi} &=&0\\
\bra{\psi}\bullet \bra{\psi} A^{(2)} \ket{\psi}\bullet \ket{\psi} &=&0\\
\bra{\psi}\bullet \bra{\psi}\bullet \bra{\psi}  A^{(3)} \ket{\psi}\bullet 
\ket{\psi}\bullet \ket{\psi}  &=&0\\
&\vdots&
\eeqa
Before we go to higher spin, we restrict our focus on multipartite states of
qubits or spins 1/2. There, the local Hilbert space is two-dimensional 
${\cal H}_i=\CC^2=:\mathfrak{h}$ for all $i$. We will need the Pauli matrices
\beq\label{def:Paulis}
\sigma_0:=\id_2=\Matrix{cc}{1&0\\0&1}\, ,\; 
\sigma_1:=\sigma_x=\Matrix{cc}{0&1\\1&0}\, ,\; 
\sigma_2:=\sigma_y=\Matrix{cc}{0&-i\\ i&0}\, ,\; 
\sigma_3:=\sigma_z=\Matrix{cc}{1&0\\0&-1}\, ,
\eeq
and highlight that any tensor product
$f(\{\sigma_\mu\}):=\sigma_{\mu_1}\otimes\dots\otimes\sigma_{\mu_n}$ 
with an odd number $N_y$ of $\sigma_y$ is a comb.
In particular, 
$
       L^{(1)}_{1/2}:=\sigma_y 
$
is the unique comb of order 1. Throughout the work the Einstein sum convention is used.
The unique comb of order 2,
which is orthogonal with respect to the Hilbert-Schmidt scalar product 
to the trivial one $\sigma_y\bullet\sigma_y$, is then
$$
L^{(2)}_{1/2}:=\sigma_\mu\bullet\sigma^\mu=
\sum_{\mu,\nu=0}^3 g^{\mu,\nu}\sigma_\mu
		\bullet\sigma_\nu\; ,
$$
with
$$ 
g^{\mu,\nu}=g_{\mu,\nu}=\Matrix{cccc}{
-1&0&0&0\\
0&1&0&0\\
0&0&0&0\\
0&0&0&1}\; .
$$
Finally, the comb of third order is
\beq
L^{(3)}_{1/2}:=\eps_{ijk}\tau_i\bullet\tau_j\bullet\tau_k\; ,
\eeq
where $\tau_1=\sigma_0$, $\tau_2=\sigma_1$, $\tau_3=\sigma_3$.
This comb completes the set of local invariant operators 
for spin $1/2$ in that it closes the algebra of antilinear operators~\cite{DoOs08}. 

Filter invariants for $n$ qubits are then obtained as 
antilinear expectation values of filter operators;
the latter are constructed from combs as to have vanishing expectation value 
for arbitrary product states (hence, they are zero for any bipartite state).
We will use the word {\em filter}\/ for both the filter operator and its filter 
invariant.
Filters for qubits have been constructed in~\cite{Wootters98,OS04,OS05,DoOs08}.

\section{Local antilinear operators for higher spin}
\label{higherSpin}

An elementary polynomial $SL$-invariant is the determinant for two qudits
(see \cite{Cereceda03} for spin 1) 
\beqa
\ket{\psi_{AB}}&=&\sum_{i,j=0}^S \psi_{ij} \ket{ij}\\
\det M_{\psi_{AB}}&=&\left|
\begin{array}{ccc}
\psi_{00}&\dots&\psi_{0S}\\
\vdots & \ddots & \vdots \\
\psi_{S0}& \dots & \psi_{SS}
\end{array}
\right|\ .
\eeqa 
Its polynomial degree is $2S+1$ and so the lowest polynomial degree 
we can expect for an invariant operator to exist, 
is $2S+1$ for spin $S$.
Because any expectation value of every operator leads to an even degree
in the wave function, there are two cases to distinguish. 
First, the case of even dimension of the 
local Hilbert space, where the determinant could be expressible in terms 
of local invariant operators, as is the case for each odd half-integer spin 
as e.g. qubits. Second, there is the case of odd dimension of the local Hilbert space,
which is given for each integer spin. 
The first occurrence of this latter scenario is for spin 1, which we will 
discuss next.

\subsection{The case of Spin 1}
Spin 1 is the first example for an odd dimension of the local Hilbert space.
The invariance group for this case is the $SL(3)$.
The Gell-Mann matrices are generators on the algebra level
\beqa
\lambda_0=\Matrix{ccc}{1&0&0\\0&1&0\\0&0&1}& 
\lambda_1=\Matrix{ccc}{0&1&0\\1&0&0\\0&0&0}&
\lambda_2=\Matrix{ccc}{0&-i&0\\i&0&0\\0&0&0}\\
\lambda_3=\Matrix{ccc}{1&0&0\\0&-1&0\\0&0&0}& 
\lambda_4=\Matrix{ccc}{0&0&1\\0&0&0\\1&0&0}&
\lambda_5=\Matrix{ccc}{0&0&-i\\0&0&0\\i&0&0}\\
\lambda_6=\Matrix{ccc}{0&0&0\\0&0&1\\0&1&0}&
\lambda_7=\Matrix{ccc}{0&0&0\\0&0&-i\\0&i&0}&
\lambda_8=\Matrix{ccc}{1&0&0\\0&1&0\\0&0&-2}\; ,
\eeqa 
including the identity matrix $\lambda_0$.
The $SL$-invariant of lowest polynomial degree is the 
determinant\cite{Cereceda03,GourGconcurrence05}, 
which is of degree 3.
In order to obtain an even polynomial degree, we take its square, which is of degree $6$, and therefore is 
achievable as an expectation value of some 
local invariant operator.
The invariant operator we obtain is therefore a comb of order 3, and is given as follows
\beq
 L^{(3)}_{1}:=i \eps_{ijk}\tau_i\bullet\tau_j\bullet\tau_k 
\eeq
where
\beq
\tau_1=\lambda_2\; ;\quad \tau_2=\lambda_5\; ;\quad \tau_3=\lambda_7
\; .
\eeq
It is worth noticing that the operators $\lambda_i$ for $i=2,5,7$ are 
precisely the matrices $\sigma_y^{(1,2)}$,  $\sigma_y^{(1,3)}$, and   
$\sigma_y^{(2,3)}$, respectively. Hence, they are combs of order 1
that however transform to a non-trivial orbit under $SL(3)$ operations\cite{OS09}.  
The square of the determinant 
\beqa
T_2^{(1)}[\psi_{12}]={\det}^2 M_{\psi_{12}}&=&\left | 
\begin{array}{ccc}
\psi_{1,1}&\psi_{1,0}&\psi_{1,-1}\\
\psi_{0,1}&\psi_{0,0}&\psi_{0,-1}\\
\psi_{-1,1}&\psi_{-1,0}&\psi_{-1,-1}
\end{array}
\right|^2\nonumber\\
&=& \left(\psi_{1,1}\psi_{0,0}\psi_{-1,-1}+\psi_{1,0}\psi_{0,-1}\psi_{-1,1}+\psi_{1,-1}\psi_{0,1}\psi_{-1,0}\right .\nonumber\\
&& \left. -\psi_{-1,1}\psi_{0,0}\psi_{1,-1}-\psi_{-1,0}\psi_{0,-1}\psi_{1,1}-\psi_{-1,-1}\psi_{0,1}\psi_{1,0}\right)^2
\eeqa
is obtained by
\beq
-\frac{1}{48}\eps_{i_1j_1k_1}\eps_{i_2j_2k_2}\bra{\Psi^*_{12}}\bullet \bra{\Psi^*_{12}}\bullet \bra{\Psi^*_{12}}
(\tau_{i_1}\otimes\tau_{i_2}) \bullet (\tau_{j_1}\otimes\tau_{j_2})
\bullet (\tau_{k_1}\otimes \tau_{k_2})
\ket{\Psi_{12}}\bullet \ket{\Psi_{12}}\bullet \ket{\Psi_{12}}\; .
\eeq

Since the procedure to obtain higher order combs 
is not described in detail in \cite{DoOs08}, 
it is worth to make some remarks here.
We consider the comb property of order $n$
\beq\label{combrel}
\bra{\psi}\bullet \bra{\psi}\bullet\dots\bullet \bra{\psi}  A^{(n)} \ket{\psi}\bullet 
\ket{\psi}\bullet\dots \bullet \ket{\psi}  =0\; .
\eeq
This condition surely does not depend on how we reorder the local states $\ket{\psi}$
on either side. Therefore, it is invariant under the full symmetric group, 
$S_n$, which means that also
\beq\label{combrel2}
\bra{\psi}\bullet \bra{\psi}\bullet\dots\bullet \bra{\psi}\Pi'  A^{(n)}\Pi \ket{\psi}\bullet 
\ket{\psi}\bullet\dots \bullet \ket{\psi}  =0
\eeq
for $\Pi,\Pi'\in S_n$. This means that if $A^{(n)}$ is a comb of order
$n$ (satisfying hence the property \eqref{combrel}), also the operator
$\Pi' A^{(n)} \Pi$ satisfies the comb relation (Eq. \eqref{combrel2}).
It therefore is a comb of order $n$ as well.\\
In what follows, we introduce the symbol $\circ$, which is used in the same way as $\bullet$ 
for highlighting the different nature of tensor products in the sequel. It symbolizes 
tensor products of the same local state, but where the permutation operator acts on. 

In order to give a simple example, we look at the comb of first order
for qubits, $\sigma_y$. The corresponding comb of second order is
$\sigma_y\circ\sigma_y$, and hence another comb of the same order $2$ is
$(\sigma_y\circ\sigma_y) \P_2=-\frac{1}{2}(\sigma_\mu\circ\sigma^\mu
-\sigma_y\circ\sigma_y)$ as reported in \cite{DoOs08}.
Next, the procedure is to render both operators orthogonal in the trace
norm and we have an orthogonal comb of order $2$. This is a straight forward
method for constructing combs of any order for $d$-dimensional qudits.
I will apply this line of thought to spin $1$.

The 2nd comb of order six will emerge from a multiplication of the trivial comb
$$
\eps_{ijk}\eps_{lmn}(\tau_i\circ
\tau_l)\bullet(\tau_j\circ \tau_m)\bullet(\tau_k\circ \tau_n)
$$
with the 
permutation operator
\beq
\P_3=\frac{1}{3} \id + \frac{1}{2} \sum_{i=1}^7 \lambda_i\circ \lambda_i
+\frac{1}{6} \lambda_8\circ \lambda_8
\eeq
as
\beq
{\cal L}^{(6)}_{1}:=-\eps_{ijk}\eps_{lmn}(\tau_i\circ
\tau_l)\bullet(\tau_j\circ \tau_m)\bullet(\tau_k\circ \tau_n)
\P_3\bullet\P_3\bullet\P_3 \; .
\eeq

We at first have to determine the nine operators 
$Q_{il}=(\tau_i\circ \tau_l)\P_3$,
which are obtained in the following form
\beqa
O_{11} &=& \frac{1}{2}(\lambda_2\circ \lambda_2 - \lambda_1\circ \lambda_1 - \lambda_3\circ \lambda_3 ) \nonumber\\
&&+ \frac{1}{18} (2 \lambda_0 + \lambda_8)\circ (2 \lambda_0 + \lambda_8)=:\xi_{11;\mu}\circ\xi_{11}^\mu\\
O_{12} &=&\frac{1}{12}((2 \lambda_0 + \lambda_8+ 3 \lambda_3)\circ (\lambda_6-i\lambda_7)+(\lambda_6+i \lambda_7)\circ (2 \lambda_0 + \lambda_8+ 3 \lambda_3))\nonumber \\
&&-\frac{1}{4}((\lambda_1-i \lambda_2)\circ( \lambda_4-i \lambda_5)+(\lambda_4+i\lambda_5)\circ (\lambda_1+i\lambda_2) )=:\xi_{12;\mu}\circ\xi_{12}^\mu\\
O_{13} &=&   \frac{1}{4}(( \lambda_1+i\lambda_2)\circ( \lambda_6-i\lambda_7) +
(\lambda_6+i\lambda_7)\circ (\lambda_1-i\lambda_2)) \nonumber\\
&&-\frac{1}{12}((2 \lambda_0 + \lambda_8- 3 \lambda_3)\circ (\lambda_4-i\lambda_5)+(\lambda_4+i \lambda_5)\circ (2 \lambda_0 + \lambda_8- 3 \lambda_3))=:\xi_{13;\mu}\circ\xi_{13}^\mu \\
O_{21} &=&\frac{1}{12}( (\lambda_6-i\lambda_7)\circ(2 \lambda_0 + \lambda_8+ 3 \lambda_3)+(2 \lambda_0 + \lambda_8+ 3 \lambda_3)\circ (\lambda_6+i \lambda_7))\nonumber\\
&&-\frac{1}{4}((\lambda_4-i\lambda_5)\circ (\lambda_1 -i\lambda_2)+
 (\lambda_1+i\lambda_2)\circ(\lambda_4+i\lambda_5) )=:\xi_{21;\mu}\circ\xi_{21}^\mu \\
O_{22} &=& \frac{1}{2}(\lambda_5\circ\lambda_5- \lambda_4\circ \lambda_4 -\frac{1}{4}(\lambda_3 + \lambda_8)\circ (\lambda_3 + \lambda_8) ) \nonumber\\
&&+ \frac{1}{72} (4 \lambda_0 +3 \lambda_3 -\lambda_8)\circ (4 \lambda_0 +3\lambda_3 - \lambda_8)=:\xi_{22;\mu}\circ\xi_{22}^\mu \\
O_{23} &=&-\frac{1}{4}( (\lambda_6-i\lambda_7)\circ( \lambda_4-i\lambda_5) +
(\lambda_4+i\lambda_5) \circ( \lambda_6+i\lambda_7) )\nonumber \\
&&+\frac{1}{6}(( \lambda_0 - \lambda_8)\circ (\lambda_1-i\lambda_2)+(\lambda_1+i \lambda_2)\circ (\lambda_0 - \lambda_8))=:\xi_{23;\mu}\circ\xi_{23}^\mu  
\eeqa
\beqa
O_{31} &=&  \frac{1}{4}((\lambda_6-i \lambda_7)\circ (\lambda_1 +i\lambda_2)+
 (\lambda_1-i\lambda_2) \circ(\lambda_6+i\lambda_7) ) \nonumber\\
&&-\frac{1}{12}( (\lambda_4-i\lambda_5)\circ(2 \lambda_0 + \lambda_8- 3 \lambda_3)+(2 \lambda_0 + \lambda_8- 3 \lambda_3)\circ(\lambda_4+i \lambda_5) )=:\xi_{31;\mu}\circ\xi_{31}^\mu \\
O_{32} &=& \frac{1}{4}((\lambda_4-i\lambda_5)\circ (\lambda_6-i\lambda_7) +
 (\lambda_6+i\lambda_7)\circ (\lambda_4+i\lambda_5) ) \nonumber\\
&&+\frac{1}{6}((\lambda_1-i\lambda_2)\circ (2 \lambda_0 - \lambda_8)+(\lambda_0 - \lambda_8)\circ(\lambda_1+i \lambda_2) )=:\xi_{32;\mu}\circ\xi_{32}^\mu \\
O_{33} &=& \frac{1}{2}(\lambda_7\circ\lambda_7-\lambda_6\circ \lambda_6 -\frac{1}{4}(\lambda_3 - \lambda_8)\circ (\lambda_3 - \lambda_8 )) \nonumber\\
&&+ \frac{1}{72} (4 \lambda_0 -3 \lambda_3 - \lambda_8)\circ (4 \lambda_0 -3 \lambda_3- \lambda_8)=:\xi_{33;\mu}\circ\xi_{33}^\mu \; .
\eeqa
It is to be mentioned that $O_{jk}=O_{kj}^T$, where $T$ means the transposition.
This is due to the relation $Q_{il}=(\tau_i\circ \tau_l)\P_3=\P_3(\tau_l\circ \tau_i)\P_3\P_3=
\P_3(\tau_l\circ \tau_i)$.
Every single operator $O_{ij}$ consists of precisely four contributions, 
each having a single entry in the matrix, 
two with contribution +1, and two with -1.
The contraction, indicated with upper and lower greek indices,
indicates to sum over these four elements in each of the $O_{ij}$. 
It is interesting that besides the $\sigma_y^{(i,j)}$ appearing
in the operators $O_{kk}$, the remaining part is
$\sigma_\mu^{(i,j)}\circ\sigma^{\mu;(i,j)}=\sum_{\mu=0}^3 g_\mu\; \sigma_\mu^{(i,j)}\circ\sigma^{\mu;(i,j)}$
with $g_\mu=(-1,1,0,1)$, resembling the situation
of the two-dimensional Hilbert space for spin $1/2$.

Using this, we have to consider
\beq
L^{(6)}_{1}:=-\eps_{ijk}\eps_{lmn} O_{il}\bullet O_{jm}\bullet O_{kn}=
-\sum_{i,j,k=0}^3\sum_{l,m,n=0}^3\eps_{ijk}\eps_{lmn} (\xi_{il;\mu}\circ\xi_{il}^\mu)\bullet(\xi_{jm;\nu}\circ\xi_{jm}^\nu)\bullet(\xi_{kn;\rho}\circ\xi_{kn}^\rho)
\eeq
as a comb of order $6$.
This operator, however, still fails to be orthogonal to the original operator
$L_1^{(3)}\circ L_1^{(3)}$ in that we have 
$\tr (L_1^{(3)}\circ L_1^{(3)}) L_1^{(6)}=31104$.
Since $\tr L_1^{(3)}\circ L_1^{(3)}= 2304$, this trace is removed in 
$L_1^{(6)}-\frac{27}{2}(L_1^{(3)}\circ L_1^{(3)})$ such that both operators 
are orthogonal.
I want to remind a fact that is also crucial here: as soon as operators as 
$\sigma_y^{(ij)}$ occur an odd number of times, its corresponding antilinear
expectation value 
$(\!(f(\sigma_\mu^{(ij)}))\!)$ 
will vanish (see for example \cite{OS04}). Therefore, 
the operator $L_1^{(6)}$ can be directly inserted without 
resorting on the orthogonal version. 
In principle, with these two operators $L_1^{(3)}$ and $L_1^{(6)}$,
we are in the position to write down SL-invariant 
entanglement measures for an arbitrary number of distinguishable particles 
of spin 1. \\
The advantage of taking the two orthogonal operators 
is the better control on the zeros of corresponding 
filter invariants (see e.g. \cite{OS05} for spin $1/2$). 

The square of the determinant\cite{Cereceda03,GourGconcurrence05}, 
which leads to a degree $6$ analogue of the concurrence for qutrits 
has been mentioned already.
An analogue of the three-tangle for three qutrits is consequently expressed by
\beqa
T_3^{(1)}[\psi_{123}]&=&\eps_{i_1j_1k_1}\eps_{l_1m_1n_1}\eps_{i_2j_2k_3}\eps_{l_2m_2n_2}\eps_{ijk}\eps_{lmn}\nonumber\\
&&\bra{\Psi^*_{123}}\bullet \bra{\Psi^*_{123}}\bullet \bra{\Psi^*_{123}}\circ
\bra{\Psi^*_{123}}\bullet \bra{\Psi^*_{123}}\bullet \bra{\Psi^*_{123}}\nonumber\\
&&\quad (\tau_{i_1}\otimes\tau_{i_2}\otimes \xi_{i l;\mu})\bullet
(\tau_{j_1}\otimes\tau_{j_2}\otimes \xi_{j m;\nu})\bullet
(\tau_{k_1}\otimes\tau_{k_2}\otimes \xi_{k n;\rho})\\
&&\qquad\qquad\qquad\qquad \circ (\tau_{l_1}\otimes\tau_{l_2}\otimes \xi_{i l}^\mu)\bullet
(\tau_{m_1}\otimes\tau_{m_2}\otimes \xi_{j m}^\nu)\bullet
(\tau_{n_1}\otimes\tau_{n_2}\otimes \xi_{k n}^\rho)\nonumber\\
&& \qquad\qquad\qquad\qquad \qquad\qquad\qquad\qquad \ket{\Psi_{123}}\bullet \ket{\Psi_{123}}\bullet \ket{\Psi_{123}}\circ
\ket{\Psi_{123}}\bullet \ket{\Psi_{123}}\bullet \ket{\Psi_{123}}\; .\nonumber
\eeqa
It is a three-qutrit filter. 
I have not checked for invariant operators existing besides this 
(see Ref.~\onlinecite{DoOs08} for qubits). 
It is possible that one will even have to look for degree $6$, 
where the permutations are set in a distinct way. 
This will be left for future work.

\subsection{The case of Spin 3/2}

The next higher case of half integer spin is spin-$3/2$, corresponding to an
even dimension of the local Hilbert space, which is $4$. Here, it is sufficient 
to prepare the determinant itself with expectation values of antilinear operators.
The underlying group is $SL(4)$, 
whose generators, including the identity matrix, read
\beq
\begin{array}{cccc}
\lambda_0=\Matrix{cccc}{1&0&0&0\\0&1&0&0\\0&0&1&0\\0&0&0&1}& 
\lambda_1=\Matrix{cccc}{0&1&0&0\\1&0&0&0\\0&0&0&0\\0&0&0&0}&
\lambda_2=\Matrix{cccc}{0&-i&0&0\\i&0&0&0\\0&0&0&0\\0&0&0&0}&
\lambda_3=\Matrix{cccc}{0&0&1&0\\0&0&0&0\\1&0&0&0\\0&0&0&0}
\\
\lambda_4=\Matrix{cccc}{0&0&-i&0\\0&0&0&0\\i&0&0&0\\0&0&0&0}&
\lambda_5=\Matrix{cccc}{0&0&0&1\\0&0&0&0\\0&0&0&0\\1&0&0&0}&
\lambda_6=\Matrix{cccc}{0&0&0&-i\\0&0&0&0\\0&0&0&0\\i&0&0&0}&
\lambda_7=\Matrix{cccc}{0&0&0&0\\0&0&1&0\\0&1&0&0\\0&0&0&0}
\\
\lambda_8=\Matrix{cccc}{0&0&0&0\\0&0&-i&0\\0&i&0&0\\0&0&0&0}&
\lambda_9=\Matrix{cccc}{0&0&0&0\\0&0&0&1\\0&0&0&0\\0&1&0&0}&
\lambda_{10}=\Matrix{cccc}{0&0&0&0\\0&0&0&-i\\0&0&0&0\\0&i&0&0}&
\lambda_{11}=\Matrix{cccc}{0&0&0&0\\0&0&0&0\\0&0&0&1\\0&0&1&0}
\\
\lambda_{12}=\Matrix{cccc}{0&0&0&0\\0&0&0&0\\0&0&0&-i\\0&0&i&0}&
\lambda_{13}=\Matrix{cccc}{1&0&0&0\\0&-1&0&0\\0&0&0&0\\0&0&0&0}&
\lambda_{14}=\Matrix{cccc}{0&0&0&0\\0&0&0&0\\0&0&1&0\\0&0&0&-1}&
\lambda_{15}=\Matrix{cccc}{1&0&0&0\\0&1&0&0\\0&0&-1&0\\0&0&0&-1}\; ,
\end{array}
\eeq 
The comb of lowest order $2$ is
\beq
L^{(2)}_{3/2}:=\sum_{i=1}^6 (-1)^{\min \{i,7-i\}}\tau_i\bullet \tau_{7-i}\; ,
\eeq
where we defined
\beq
\tau_i=\lambda_{2i}\; ;\ \mbox{ for }i=1,\dots,6
\; .
\eeq
It reproduces the determinant of two qudit states of spin-3/2 
in the following way
\beq
\frac{1}{24} \sum_{i,j=1}^6 (-1)^{\min \{i,7-i\}+\min \{j,7-j\}}
(\!(\tau_i\otimes\tau_j)\bullet (\tau_{7-i}\otimes\tau_{7-j})\!)
=\left|
\begin{array}{cccc}
\psi_{2,2}&\psi_{2,1}&\psi_{2,-1}&\psi_{2,-2}\\
\psi_{1,2}&\psi_{1,1}&\psi_{1,-1}&\psi_{1,-2}\\
\psi_{-1,2}&\psi_{-1,1}&\psi_{-1,-1}&\psi_{-1,-2}\\
\psi_{-2,2}&\psi_{-2,1}&\psi_{-2,-1}&\psi_{-2,-2}\\
\end{array}\right|
\eeq
Please notice that the operators $\tau_i$ correspond to
$\lambda_{2i}$ which are the $\sigma_y^{(k,l)}$ acting on the subspaces $k$ and $l$, 
where $(k,l)$ takes the values $(1,2),(1,3),(1,4),(2,3),(2,4)$ 
and $(3,4)$ of the four-dimensional local space.

The second comb is of order $4$ (hence, it is related to polynomial degree $8$), 
and is obtained by means of the permutation operator 
\beq
\P_4=\frac{1}{4} \id + \frac{1}{2} \sum_{i=1}^{14} \lambda_i \circ \lambda_i
+ \frac{1}{4} \lambda_{15}\circ \lambda_{15}
\eeq
in the following way
\beq
L^{(4)}_{3/2}:=\sum_{i,j=1}^6 (-1)^{\min \{i,7-i\}+\min \{j,7-j\}}(\tau_i\circ\tau_j)\bullet (\tau_{7-i}\circ \tau_{7-j})\P_4\bullet\P_4\; .
\eeq
Therefore, the operators $O_{ij}=(\tau_i\circ\tau_j)\P_4$ are relevant, 
which are presented in the appendix.
Again, we find $O_{jk}=O_{kj}^T$. 
Also here it is striking that, similar to the previous case, 
the operators $O_{jk}$ have only four contributions, 
corresponding to four entries (two with +1, and two with -1).
The comb of order 4 is hence
\beq
L^{(4)}_{3/2}=\sum_{i,j=1}^6(-1)^{\min \{i,7-i\}+ \min \{j,7-j\}}
O_{ij}\bullet O_{7-i,7-j}=\sum_{i,j=1}^6(-1)^{\min \{i,7-i\}+ \min \{j,7-j\}} (\xi_{ij;\mu}\circ\xi_{ij}^\mu)\bullet(\xi_{7-i,7-j;\mu}\circ\xi_{7-i,7-j}^\mu)
\eeq
This operator is again not orthogonal to the trivial one, 
$L^{(2)}_{3/2}\circ L^{(2)}_{3/2}$, in that we find 
$\tr L ^{(4)}_{3/2} (L^{(2)}_{3/2}\circ L^{(2)}_{3/2})=\frac{3}{2}$. 
Together with $tr L^{(2)}_{3/2}\circ L^{(2)}_{3/2}=9$ we obtain the orthogonal
operator $L ^{(4)}_{3/2} - \frac{1}{6} L^{(2)}_{3/2}\circ L^{(2)}_{3/2}$.
These two orthogonal combs are the counterparts of $\sigma_y$ and 
$\sigma_\mu\circ \sigma^\mu$ for qubits.
They are sufficient for the construction 
of many (most likely not of all) $SL$-invariants for spin 3/2. 
However, as for the spin-1 case, using the operator  $L ^{(4)}_{3/2}$ 
an analogue of the three-tangle for spin-3/2 can be expressed as
\beqa
T_3^{(3/2)}[\Psi_{123}]&=&\frac{1}{8}\sum_{i,j,k,l=1}^6(-1)^{\min \{i,7-i\}+\min \{j,7-j\}+\min \{k,7-k\}+\min \{l,7-l\}}\sum_{m,n=1}^6(-1)^{\min \{m,7-m\}+\min \{n,7-n\}}\nonumber\\
&&\bra{\Psi^*_{123}}\bullet \bra{\Psi^*_{123}}\circ\bra{\Psi^*_{123}}\bullet
\bra{\Psi^*_{123}}\nonumber\\
&&\quad(\tau_i\otimes\tau_j\otimes\xi_{mn;\mu})\bullet 
(\tau_{7-i}\otimes \tau_{7-j}\otimes\xi_{7-m,7-n;\nu})
\ \circ \
 (\tau_k\otimes\tau_l\otimes\xi_{mn}^\mu)\bullet 
(\tau_{7-k}\otimes \tau_{7-l}\otimes\xi_{7-m,7-n}^\nu)\nonumber\\
&& \qquad\qquad\qquad\qquad\qquad\qquad\quad \ket{\Psi_{123}}\bullet \ket{\Psi_{123}}\circ\ket{\Psi_{123}}\bullet 
\ket{\Psi_{123}}\; .\nonumber
\eeqa
The existence of further combs of higher order will be left for future work.

\section{Conclusion}\label{concls}
I have outlined a path towards the generalization of entanglement measures 
along the line pursued in \cite{Wootters98,Wong00,OS05} to systems 
with higher local dimensions i.e. for higher spin $S$. 
Therefore, the theory of local 
SL-invariant operators~\cite{OS04} has been considered and developed further.
A comb, written in terms of expectation values, has an even 
polynomial degree. The lowest possible such degree would be $2S+1$,
due to the determinant of a $(2S+1)\times(2S+1)$ matrix. This lowest
possible degree however can only 
be realized for an even dimension, hence for $S=m/2$ for odd $m$. 
In case of an odd dimension $2S+1$, i.e. for integer spin $S$,
the lowest dimension is therefore doubled as $4S+2$. 
In order to have more than a single SL-invariant, we need at least one more and linearly independent 
local antilinear operator which is possibly (but not necessarily) orthogonal to
the comb of lowest order or (multiple) tensor products of it. Two operators are
sufficient for being able to construct measures for genuine multipartite entanglement which are 
related to an entanglement monotone~\cite{OS04,OS05,DoOs08,Eltschka}.
Therefore, it is crucial to observe that 
the {\em comb condition} for a local SL-invariant operator is 
invariant with respect to the symmetric group (a fact which has been 
observed and used excessively already in \cite{DoOs08}). The symmetric group hence acts 
transitively on the space of combs of a certain order.
Using this method, I give expressions for two local and orthogonal SL-invariant operators 
for each spin, $1$ and $3/2$, 
corresponding to odd and even dimensions of the local Hilbert space, respectively. I give explicit formulae 
for analogues to the concurrence and the three-tangle for qubits.
Open questions remain. One of the next tasks would be to find out how this formalism is generalized 
to arbitrary dimension
as a straight forward generalization of the formulae given in the text do not lead to an answer. 
This has to be left for future work.
Also the question of how a complete set of local SL-invariant operators is obtained, 
has to be analyzed later on.

\acknowledgments
This work was supported by the German Research Foundation within the SFB TR12.

\appendix

\section{The operator for spin $3/2$}
The operators $O_{ij}=(\tau_i\circ\tau_j)\P_4$ are given here explicitly for local dimension $4$, or spin $3/2$.
We find them also being symmetric, $O_{jk}=O_{kj}^T$, and
similar to spin $1$
the operators $O_{jk}$ have only four contributions, 
corresponding to four entries (two with +1, and two with -1). 
\beqa
O_{11}&=&\frac{1}{2}(\lambda_2\circ \lambda_2- \lambda_1\circ \lambda_1 -\lambda_{13}\circ \lambda_{13})+
\frac{1}{8} (\lambda_0+\lambda_{15})\circ (\lambda_0+\lambda_{15})=\xi_{11;\mu}\circ\xi_{11}^\mu\\
O_{12}&=&-\frac{1}{4}((\lambda_1-i \lambda_2)\circ (\lambda_3 -i\lambda_{4})+
                     (\lambda_3+i \lambda_4)\circ (\lambda_1+i\lambda_2))\nonumber\\
&&+
\frac{1}{8} ((\lambda_0+2\lambda_{13}+\lambda_{15})\circ (\lambda_7-i\lambda_{8})+ (\lambda_7+i \lambda_8)\circ(\lambda_0+2\lambda_{13}+\lambda_{15}))
=\xi_{12;\mu}\circ\xi_{12}^\mu\\
O_{13}&=&-\frac{1}{4}((\lambda_1- i\lambda_2)\circ (\lambda_5 - i\lambda_{6})+
                     (\lambda_6+i \lambda_5)\circ (\lambda_1+i\lambda_2))\nonumber\\
&&+
\frac{1}{8} ((\lambda_0+2\lambda_{13}+\lambda_{15})\circ (\lambda_9-i\lambda_{10})+ (\lambda_9+i \lambda_{10})\circ(\lambda_0+2\lambda_{13}+\lambda_{15}))
=\xi_{13;\mu}\circ\xi_{13}^\mu\\
O_{14}&=&\frac{1}{4}((\lambda_1+i \lambda_2)\circ (\lambda_7 -i\lambda_{8})+
                     (\lambda_7+i \lambda_8)\circ (\lambda_1 -i\lambda_2))\nonumber\\
&&-
\frac{1}{8} ((\lambda_0-2\lambda_{13}+\lambda_{15})\circ (\lambda_3-i\lambda_{4})+ (\lambda_3+i \lambda_4)\circ(\lambda_0-2\lambda_{13}+\lambda_{15}))
=\xi_{14;\mu}\circ\xi_{14}^\mu\\
O_{15}&=&\frac{1}{4}((\lambda_1+i \lambda_2)\circ (\lambda_9 -i\lambda_{10})+
                     (\lambda_9+i \lambda_{10})\circ (\lambda_1-i\lambda_2))\nonumber\\
&&-
\frac{1}{8} ((\lambda_0-2\lambda_{13}+\lambda_{15})\circ (\lambda_5-i\lambda_{6})+ (\lambda_5+i \lambda_6)\circ(\lambda_0-2\lambda_{13}+\lambda_{15}))
=\xi_{15;\mu}\circ\xi_{15}^\mu\\
O_{16}&=&\frac{1}{4}((\lambda_9+ i \lambda_{10})\circ (\lambda_3 -i\lambda_{4})+
                     (\lambda_3+i \lambda_4)\circ (\lambda_9-i\lambda_{10}))\nonumber\\
&&-
\frac{1}{4} ((\lambda_5+i\lambda_{6})\circ (\lambda_7-i\lambda_{8})+ (\lambda_7+i \lambda_8)\circ(\lambda_5-i\lambda_{6}))
=\xi_{16;\mu}\circ\xi_{16}^\mu
\eeqa
\beqa
%O_{21}&=&-\frac{1}{4}((\lambda_1+i \lambda_2)\circ (\lambda_3 +i\lambda_{4})+
%                     (\lambda_3-i \lambda_4)\circ (\lambda_1-i\lambda_2))\nonumber\\
%&&+
%\frac{1}{8} ((\lambda_0+2\lambda_{13}+\lambda_{15})\circ (\lambda_7+i\lambda_{8})+ (\lambda_7-i \lambda_8)\circ(\lambda_0+2\lambda_{13}+\lambda_{15}))
%=\xi_{21;\mu}\circ\xi_{21}^\mu\\
O_{22}&=&\frac{1}{2}(\lambda_4\circ \lambda_4- \lambda_3\circ \lambda_3) \nonumber\\
&&-\frac{1}{8} ((\lambda_{13}-\lambda_{14}+\lambda_{15})\circ (\lambda_{13}-\lambda_{14}+\lambda_{15})+
(\lambda_0+\lambda_{13}+\lambda_{14})\circ (\lambda_0+\lambda_{13}+\lambda_{14}))=\xi_{22;\mu}\circ\xi_{22}^\mu\\
O_{23}&=&-\frac{1}{4}((\lambda_3- i\lambda_4)\circ (\lambda_5 - i\lambda_{6})+
                     (\lambda_5+i \lambda_6)\circ (\lambda_3+i\lambda_4))\nonumber\\
&&+
\frac{1}{8} ((\lambda_0+2\lambda_{13}+\lambda_{15})\circ (\lambda_{11}-i\lambda_{12})+ (\lambda_{11}+i \lambda_{12})\circ(\lambda_0+2\lambda_{13}+\lambda_{15}))
=\xi_{23;\mu}\circ\xi_{23}^\mu\\
O_{24}&=&-\frac{1}{4}((\lambda_3+i \lambda_4)\circ (\lambda_7 +i\lambda_{8})+
                     (\lambda_7-i \lambda_8)\circ (\lambda_3 -i\lambda_4))\nonumber\\
&&+
\frac{1}{8} ((\lambda_0+2\lambda_{14}-\lambda_{15})\circ (\lambda_{1}-i\lambda_{2})+ (\lambda_{1}+i \lambda_{2})\circ(\lambda_0+2\lambda_{14}-\lambda_{15}))
=\xi_{24;\mu}\circ\xi_{24}^\mu\\
O_{25}&=&-\frac{1}{4}((\lambda_7-i \lambda_8)\circ (\lambda_5 -i\lambda_{6})+
                     (\lambda_5+i \lambda_{6})\circ (\lambda_7+i\lambda_8))\nonumber\\
&&+
\frac{1}{4} ((\lambda_{11}+i\lambda_{12})\circ (\lambda_1-i\lambda_{2})+ (\lambda_1+i \lambda_2)\circ(\lambda_{11}-i\lambda_{12}))
=\xi_{25;\mu}\circ\xi_{25}^\mu\\
O_{26}&=&\frac{1}{4}((\lambda_3+ \lambda_{4})\circ (\lambda_{11} -i\lambda_{12})+
                  (\lambda_{11}+i \lambda_{12})\circ (\lambda_3-i\lambda_{4}))\nonumber\\
&&-
\frac{1}{8} ((\lambda_5+i\lambda_{6})\circ (\lambda_0+2\lambda_{14}-\lambda_{15})+ (\lambda_0+2\lambda_{14}-\lambda_{15})\circ(\lambda_5-i\lambda_{6}))
=\xi_{26;\mu}\circ\xi_{26}^\mu
\eeqa
\beqa
%O_{31}&=&-\frac{1}{4}((\lambda_5- i\lambda_6)\circ (\lambda_1 - i\lambda_{2})+
%                     (\lambda_1+i \lambda_2)\circ (\lambda_5+i\lambda_6))\nonumber\\
%&&+
%\frac{1}{8} ((\lambda_0+2\lambda_{13}+\lambda_{15})\circ (\lambda_9+i\lambda_{10})+ (\lambda_9-i \lambda_{10})\circ(\lambda_0+2\lambda_{13}+\lambda_{15}))
%=\xi_{31;\mu}\circ\xi_{31}^\mu\\
%O_{32}&=&-\frac{1}{4}((\lambda_5- i\lambda_6)\circ (\lambda_3 - i\lambda_{4})+
%                     (\lambda_3+i \lambda_4)\circ (\lambda_5+i\lambda_6))\nonumber\\
%&&+
%\frac{1}{8} ((\lambda_0+2\lambda_{13}+\lambda_{15})\circ (\lambda_{11}+i\lambda_{12})+ (\lambda_{11}-i \lambda_{12})\circ(\lambda_0+2\lambda_{13}+\lambda_{15}))
%=\xi_{32;\mu}\circ\xi_{32}^\mu\\
O_{33}&=&\frac{1}{2}(\lambda_6\circ \lambda_6- \lambda_5\circ \lambda_5) \nonumber\\
&&-\frac{1}{8} ((\lambda_{13}+\lambda_{14}+\lambda_{15})\circ (\lambda_{13}+\lambda_{14}+\lambda_{15})-
(\lambda_0+\lambda_{13}-\lambda_{14})\circ (\lambda_0+\lambda_{13}-\lambda_{14}))=\xi_{33;\mu}\circ\xi_{33}^\mu\\
O_{34}&=&-\frac{1}{4}((\lambda_9-i \lambda_{10})\circ (\lambda_3 -i\lambda_{4})+
                     (\lambda_3+i \lambda_4)\circ (\lambda_9 +i\lambda_{10}))\nonumber\\
&&+
\frac{1}{4} ((\lambda_{11}-i\lambda_{12})\circ (\lambda_{1}-i\lambda_{2})+ (\lambda_{1}+i \lambda_{2})\circ(\lambda_{11}+i \lambda_{12}))
=\xi_{34;\mu}\circ\xi_{34}^\mu\\
O_{35}&=&-\frac{1}{4}((\lambda_5+i \lambda_6)\circ (\lambda_9 +i\lambda_{10})+
                     (\lambda_9-i \lambda_{10})\circ (\lambda_5-i\lambda_6))\nonumber\\
&&+
\frac{1}{8} ((\lambda_{1}+i\lambda_{2})\circ (\lambda_0-2\lambda_{14}-\lambda_{15})+ (\lambda_0-2\lambda_{14}-\lambda_{15})\circ(\lambda_{1}-i \lambda_{2}))
=\xi_{35;\mu}\circ\xi_{35}^\mu\\
O_{36}&=&-\frac{1}{4}((\lambda_5+i \lambda_{6})\circ (\lambda_{11} +i\lambda_{12})+
                  (\lambda_{11}-i \lambda_{12})\circ (\lambda_5-i\lambda_{6}))\nonumber\\
&&+
\frac{1}{8} ((\lambda_3+i\lambda_{4})\circ (\lambda_0-2\lambda_{14}-\lambda_{15})+ (\lambda_0-2\lambda_{14}-\lambda_{15})\circ(\lambda_3-i\lambda_{4}))
=\xi_{36;\mu}\circ\xi_{36}^\mu
\eeqa
\beqa
%O_{41}&=&\frac{1}{4}((\lambda_7-i \lambda_8)\circ (\lambda_1 +i\lambda_{2})+
%                     (\lambda_1-i \lambda_2)\circ (\lambda_7 +i\lambda_8))\nonumber\\
%&&-
%\frac{1}{8} ((\lambda_0-2\lambda_{13}+\lambda_{15})\circ (\lambda_3+i\lambda_{4})+ (\lambda_3-i \lambda_4)\circ(\lambda_0-2\lambda_{13}+\lambda_{15}))
%=\xi_{41;\mu}\circ\xi_{41}^\mu\\
%O_{42}&=&-\frac{1}{4}((\lambda_7+i \lambda_8)\circ (\lambda_3 +i\lambda_{4})+
%                     (\lambda_3-i \lambda_4)\circ (\lambda_7 -i\lambda_8))\nonumber\\
%&&+
%\frac{1}{8} ((\lambda_0+2\lambda_{14}-\lambda_{15})\circ (\lambda_{1}+i\lambda_{2})+ (\lambda_{1}-i \lambda_{2})\circ(\lambda_0+2\lambda_{14}-\lambda_{15}))
%=\xi_{42;\mu}\circ\xi_{42}^\mu\\
%O_{43}&=&-\frac{1}{4}((\lambda_3-i \lambda_{4})\circ (\lambda_9 -i\lambda_{10})+
%                     (\lambda_9+i \lambda_{10})\circ (\lambda_3 +i\lambda_{4}))\nonumber\\
%&&+
%\frac{1}{4} ((\lambda_{1}-i\lambda_{2})\circ (\lambda_{11}-i\lambda_{12})
%           + (\lambda_{11}+i \lambda_{12})\circ(\lambda_{1}+i \lambda_{2}))
%=\xi_{43;\mu}\circ\xi_{43}^\mu\\
O_{44}&=&\frac{1}{2}(\lambda_8\circ \lambda_8- \lambda_7\circ \lambda_7) \nonumber\\
&&-\frac{1}{8} ((\lambda_{13}+\lambda_{14}-\lambda_{15})\circ (\lambda_{13}+\lambda_{14}-\lambda_{15})-
(\lambda_0-\lambda_{13}+\lambda_{14})\circ (\lambda_0-\lambda_{13}+\lambda_{14}))=\xi_{44;\mu}\circ\xi_{44}^\mu\\
O_{45}&=&-\frac{1}{4}((\lambda_7-i \lambda_8)\circ (\lambda_9 -i\lambda_{10})+
                     (\lambda_9+i \lambda_{10})\circ (\lambda_7+i\lambda_8))\nonumber\\
&&+
\frac{1}{8} ((\lambda_0-2\lambda_{13}+\lambda_{15})\circ (\lambda_{11}-i\lambda_{12})+ (\lambda_{11}+i \lambda_{12})\circ(\lambda_0-2\lambda_{13}+\lambda_{15}))
=\xi_{45;\mu}\circ\xi_{45}^\mu\\
O_{46}&=&\frac{1}{4}((\lambda_7+i \lambda_{8})\circ (\lambda_{11} -i\lambda_{12})+
                  (\lambda_{11}+i \lambda_{12})\circ (\lambda_7-i\lambda_{8}))\nonumber\\
&&-
\frac{1}{8} ((\lambda_9+i\lambda_{10})\circ (\lambda_0+\lambda_{14}-\lambda_{15})+ (\lambda_0+\lambda_{14}-\lambda_{15})\circ(\lambda_9-i\lambda_{10}))
=\xi_{46\;\mu}\circ\xi_{46}^\mu
\eeqa
\beqa
%O_{51}&=&\frac{1}{4}((\lambda_9-i \lambda_{10})\circ (\lambda_1 +i\lambda_{2})+
%                     (\lambda_1-i \lambda_{2})\circ (\lambda_9+i\lambda_{10}))\nonumber\\
%&&-
%\frac{1}{8} ((\lambda_0-2\lambda_{13}+\lambda_{15})\circ (\lambda_5+i\lambda_{6})+ (\lambda_5-i \lambda_6)\circ(\lambda_0-2\lambda_{13}+\lambda_{15}))
%=\xi_{51;\mu}\circ\xi_{51}^\mu\\
%O_{52}&=&-\frac{1}{4}((\lambda_5-i \lambda_6)\circ (\lambda_7 -i\lambda_{8})+
%                     (\lambda_7+i \lambda_{8})\circ (\lambda_5+i\lambda_6))\nonumber\\
%&&+
%\frac{1}{4} ((\lambda_{1}-i\lambda_{2})\circ (\lambda_{11}+i\lambda_{12})+ (\lambda_{11}-i \lambda_{12})\circ(\lambda_{1}+i\lambda_{2}))
%=\xi_{52;\mu}\circ\xi_{52}^\mu\\
%O_{53}&=&-\frac{1}{4}((\lambda_9+i \lambda_{10})\circ (\lambda_5 +i\lambda_{6})+
%                     (\lambda_5-i \lambda_{6})\circ (\lambda_9-i\lambda_{10}))\nonumber\\
%&&+
%\frac{1}{8} ((\lambda_{1}-i\lambda_{2})\circ (\lambda_0-2\lambda_{14}-\lambda_{15})+ (\lambda_0-2\lambda_{14}-\lambda_{15})\circ(\lambda_{1}+i \lambda_{2}))
%=\xi_{53;\mu}\circ\xi_{53}^\mu\\
%O_{54}&=&-\frac{1}{4}((\lambda_9-i \lambda_{10})\circ (\lambda_7 -i\lambda_{8})+
%                     (\lambda_7+i \lambda_{8})\circ (\lambda_9+i\lambda_{10}))\nonumber\\
%&&+
%\frac{1}{8} ((\lambda_0-2\lambda_{13}+\lambda_{15})\circ (\lambda_{11}+i\lambda_{12})+ (\lambda_{11}-i \lambda_{12})\circ(\lambda_0-2\lambda_{13}+\lambda_{15}))
%=\xi_{54;\mu}\circ\xi_{54}^\mu\\
O_{55}&=&\frac{1}{2}(\lambda_{10}\circ \lambda_{10}- \lambda_9\circ \lambda_9) \nonumber\\
&&-\frac{1}{8} ((\lambda_{13}-\lambda_{14}-\lambda_{15})\circ (\lambda_{13}-\lambda_{14}-\lambda_{15})-
(\lambda_0-\lambda_{13}-\lambda_{14})\circ (\lambda_0-\lambda_{13}-\lambda_{14}))=\xi_{55;\mu}\circ\xi_{55}^\mu\\
O_{56}&=&-\frac{1}{4}((\lambda_9+ i \lambda_{10})\circ (\lambda_{11} +i\lambda_{12})+
                  (\lambda_{11}-i \lambda_{12})\circ (\lambda_9-i\lambda_{10}))\nonumber\\
&&+
\frac{1}{8} ((\lambda_7+i\lambda_{8})\circ (\lambda_0-2\lambda_{14}-\lambda_{15})+ (\lambda_0-2\lambda_{14}-\lambda_{15})\circ(\lambda_7-i\lambda_{8}))
=\xi_{56\;\mu}\circ\xi_{56}^\mu
\eeqa
\beqa
O_{66}&=&\frac{1}{2}(\lambda_{12}\circ \lambda_{12}- \lambda_{11}\circ \lambda_{11} -\lambda_{14}\circ \lambda_{14})+
\frac{1}{8} (\lambda_0-\lambda_{15})\circ (\lambda_0-\lambda_{15})=\xi_{66;\mu}\circ\xi_{66}^\mu
\eeqa
Also for spin-3/2 it is intriguing that $O_{kk}$ contains 
$\sigma_\mu^{(i,j)}\circ\sigma^{\mu;(i,j)}=\sum_{\mu=0}^3 g_\mu\; \sigma_\mu^{(i,j)}\circ\sigma^{\mu;(i,j)}$
with $g_\mu=(-1,1,0,1)$,
besides an $\sigma_y^{(i,j)}$.
This is again resembling the situation
of the two-dimensional Hilbert space for spin $1/2$.

%\bibliography{biblio,Qubits.SU} 

\end{document}